\documentclass[twoside]{ilcws07}
\usepackage[latin1]{inputenc}
\usepackage[dvips]{graphicx,epsfig,color}
\usepackage{wrapfig,rotating}
\usepackage{amssymb,amsmath,array}

\def\ifmath#1{\relax\ifmmode #1\else $#1$\fi}%
\def\chio{\ifmath{\mathchoice%
     {\displaystyle\raise.4ex\hbox{$\displaystyle\tilde\chi{^0_1}$}}%
        {\textstyle\raise.4ex\hbox{$\textstyle\tilde\chi{^0_1}$}}%
      {\scriptstyle\raise.3ex\hbox{$\scriptstyle\tilde\chi{^0_1}$}}%
{\scriptscriptstyle\raise.3ex\hbox{$\scriptscriptstyle\tilde\chi{^0_1}$}}}}

\begin{document}
\begin{titlepage}

\thispagestyle{empty}
\def\thefootnote{\fnsymbol{footnote}}       

\begin{center}
\mbox{ }

\end{center}
\begin{flushright}
\vspace* {-2.0cm}
\Large
\mbox{\hspace{9cm} EUROTeV-Report-2008-020} \\
\end{flushright}
\begin{center}
\includegraphics[width=5cm]{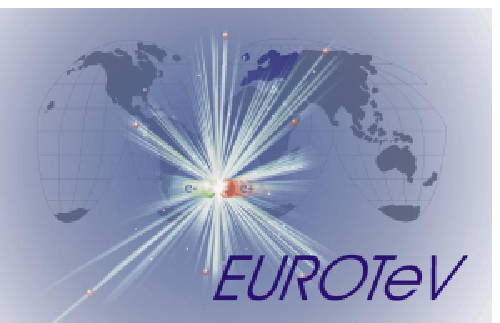}
\end{center}

\vskip 5cm
\hspace*{-2cm}
\begin{picture}(0.001,0.001)(0,0)
\put(,0){
\begin{minipage}{1.2\textwidth}
\begin{center}
\vskip 1.0cm

{\Huge\bf
Collimator R\&D
}
\vspace{3mm}

\vskip 1.5cm
{\LARGE\bf 
Andr\'e Sopczak

\bigskip

\Large Lancaster University, UK}

\vskip 2cm
{\Large \bf Abstract}
\end{center}
\end{minipage}
}
\end{picture}

\vskip 4.5cm
\hspace*{-1cm}
\begin{picture}(0.001,0.001)(0,0)
\put(,0){
\begin{minipage}{\textwidth}
\Large
\renewcommand{\baselinestretch} {1.2}
The importance of collimator R\&D with test beam facilities
is demonstrated with the example of ILC collimator studies at 
SLAC End Station A.
Related LHC collimator aspects and longer term plans are addressed.
\renewcommand{\baselinestretch} {1.}

\normalsize
\vspace{2.5cm}
\begin{center}
{\sl \large
Presented at IDTB07,
ILC Detector Test Beam workshop,\\
Fermilab, USA, 2007.
\vspace{-6cm}
}
\end{center}
\end{minipage}
}
\end{picture}
\vfill

\end{titlepage}

\clearpage
\thispagestyle{empty}
\mbox{ }
\newpage
\setcounter{page}{1}
\pagestyle{plain}

\title{Collimator R\&D}

\author{Andr\'e Sopczak\thanks{Email: andre.sopczak@cern.ch}
\vspace{.3cm}\\
 Lancaster University 
}

\maketitle

\begin{abstract}
The importance of collimator R\&D with test beam facilities
is demonstrated with the example of ILC collimator studies at 
SLAC End Station A.
Related LHC collimator aspects and longer term plans are addressed.
\end{abstract}

\section{Introduction}
Test beams have proven to be crucial for many developments
in accelerator and detector physics.
The importance of test beams for the collimator development is
discussed.
Both, accelerator and detector physics profit from collimator studies. 
On the detector side most significant is the vertex detector 
which has the smallest aperture (radius) at the interaction point. 
Vertex detector design aspects (e.g. innermost radius, 
occupancy due to background rate and required radiation tolerance)
and collimator design studies are related.
Precise collimation of the beam halo could prevent beam losses near 
the interaction region that could cause unacceptable backgrounds 
in the detector. 
The collimator design must be optimized for minimal beam disturbance 
and maximal detector protection.
Particular attention is devoted to wakefields induced by the collimators.
The optimization of the collimator shape requires simulations and test 
beam studies.
The challenges involved in the collimator design with tight apertures are 
related to wakefields that deflect the beam and increase the emittance.
Much progress has been made in recent years in the collimator development
with dedicated ILC 
studies~\cite{Woods:2005vy,Watson:2006dz,Woods:2006bs,pac07}.
At the LHC collimator design studies are crucial for the protection
of the accelerator and are a challenge owing to the large stored energy
of the beam~\cite{Assmann:2006bd}.

\section{SLAC's End Station A (ESA) Test Facility}

In 2006 and 2007 two collimator test beam studies took place each year.
The collimator wakefield measurements project is T-480.
The collaborating institutions have been 
Birmingham University, UK;
CCLRC ASTeC, UK;
CCLRC Engineering Department, UK;
CERN, Switzerland;
Manchester University, UK;
Lancaster University, UK;
DESY, Germany;
TEMF TU Darmstadt, Germany; and 
SLAC, USA.

The two sets of test beam measurements performed at the SLAC ESA 
facility in 2006 showed the world-wide cooperation
in this field of research which is expressed by
Jonathan Dorfan in SLAC Today, May 8, 2006:
``Living the "I" of the ILC It's Happening Right Here. Here at SLAC.
...
Forty physicists from 15 institutions are participating in a 
two-week run, completing today, for ILC beam tests at the 
End Station A facility.
...
To make these tests realistic and useful, the researchers need a 
beam which has the challenging bunch parameters needed for the ILC. 
There is only one place in the world where that is possible: here at 
SLAC. The unique SLAC electron beam is transported faithfully to 
End Station A, where a mature user facility, ideally suited to 
efficient and effective testing, exists.''

\subsection{Collimators for Study at ESA in 2006}

In the 2006 test beam studies at ESA eight collimator shapes,
which were produced in UK and shipped to SLAC, were tested.
These tests are aimed at measuring the wakefield kick from the collimator
structures.
At SLAC a single beam line brings primary electrons from the 
main linac to End Station A, with energies up to 28.5 GeV.
The energy spread is 0.2\% and the beam spot size about 
($x,y)=(100,600)\mu$m. The pulse repetition rate was 10~Hz.

\begin{figure}[htbp]
\vspace*{5mm}
\begin{center}
\includegraphics[width=\textwidth]{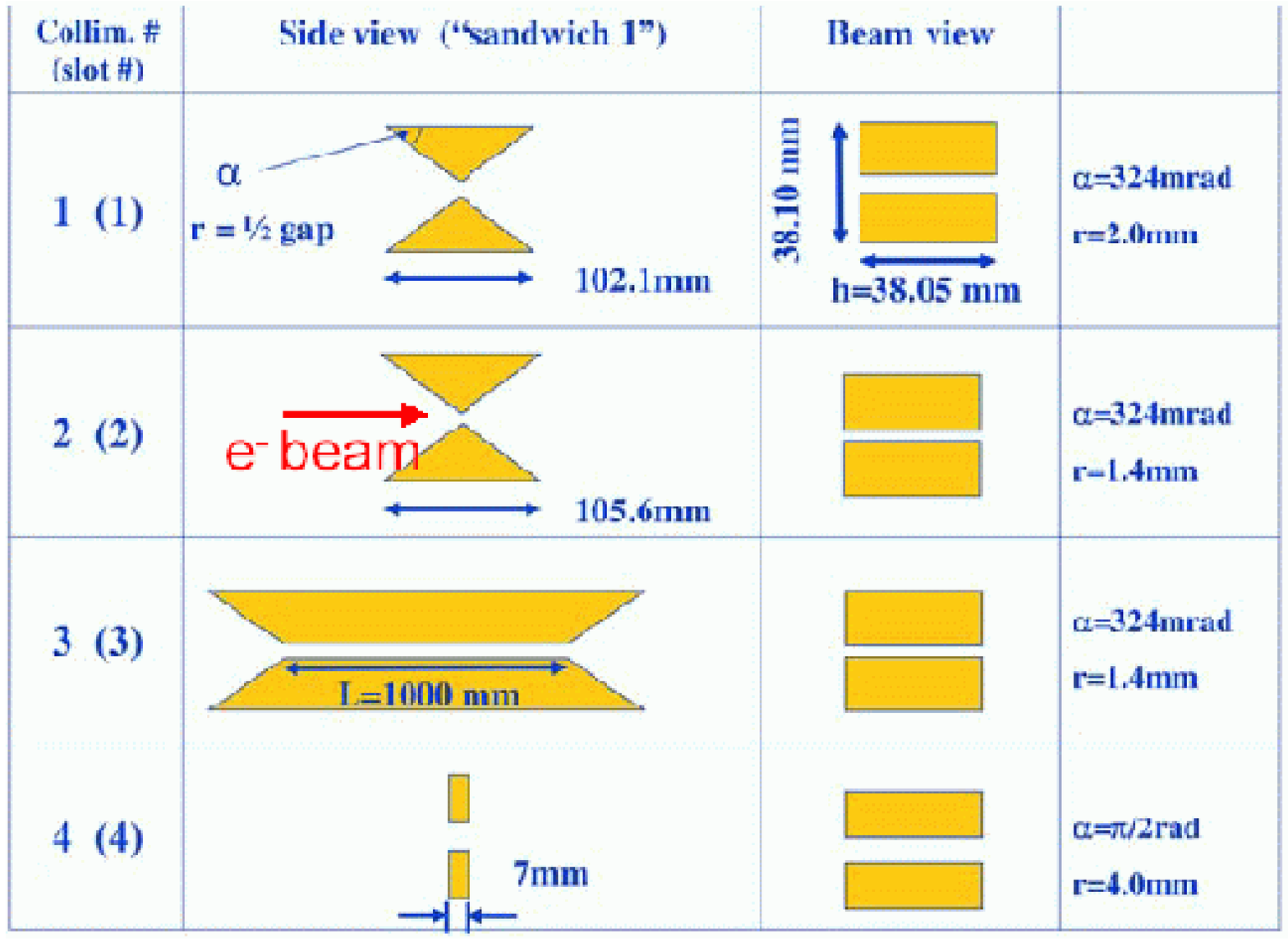}
\end{center}
\vspace*{-5mm}
\caption{Collimator designs for wakefield measurements. Sandwich 1.
\label{fig:collimator1}}
\end{figure}

\begin{figure}[htbp]
\begin{center}
\includegraphics[width=\textwidth]{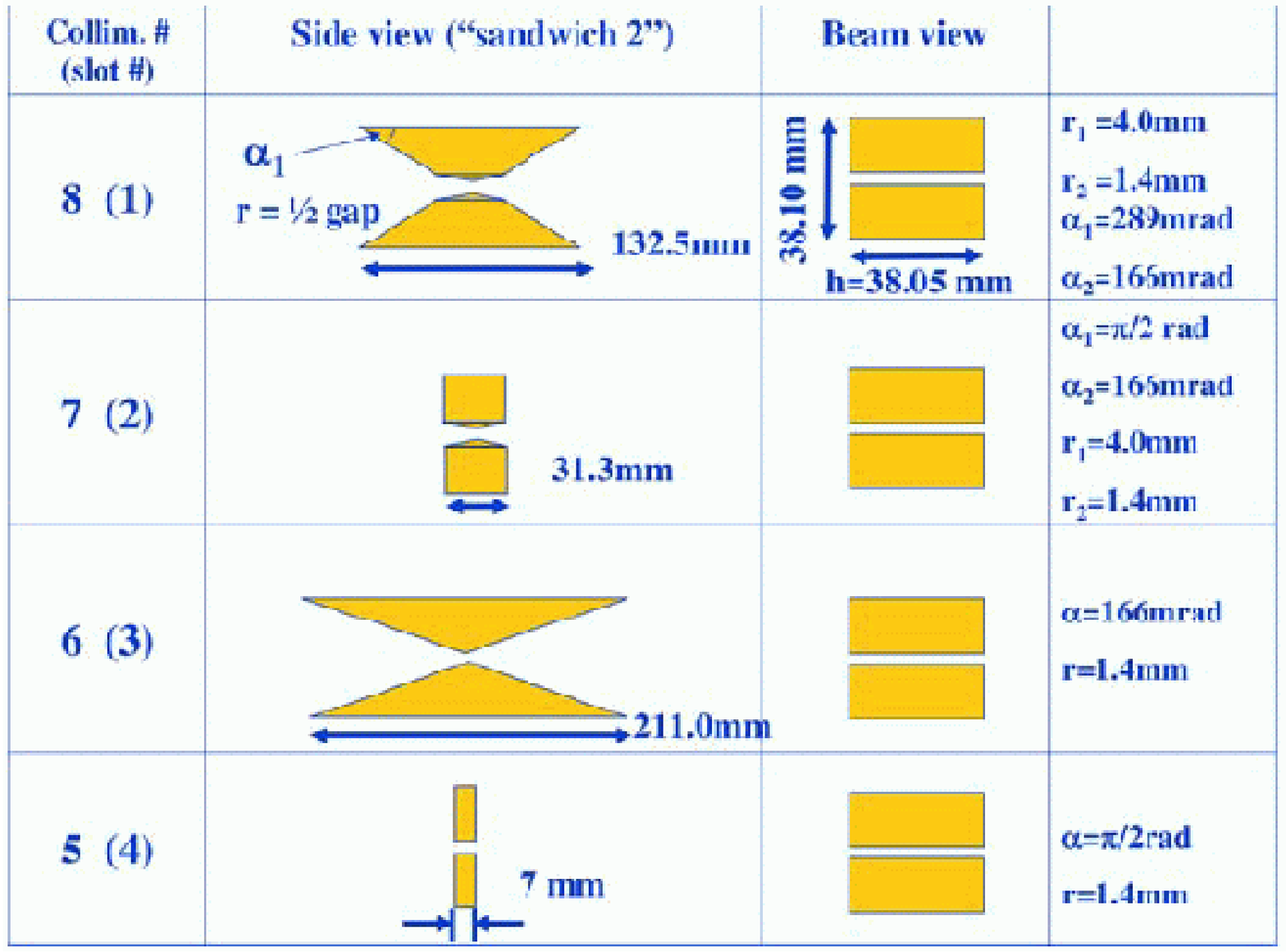}
\end{center}
\vspace*{-5mm}
\caption{Collimator designs for wakefield measurements. Sandwich 2.
\label{fig:collimator2}}
\end{figure}

\begin{figure}[htbp]
\begin{center}
\includegraphics[width=\textwidth]{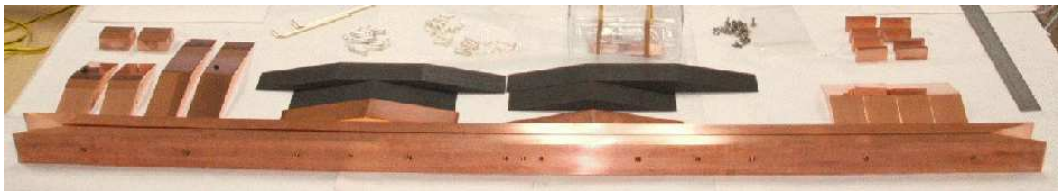}
\end{center}
\vspace*{-5mm}
\caption{Produced collimator jaws for wakefield measurements.
         The longest pair of collimators is shown in the front of the
         picture and has a length of 1m.
\label{fig:collimators_small.ps}}
\end{figure}

The eight different collimator types have been assembled 
in two sandwich structures. The schematic view
and design parameters of the collimators are shown in
Figs.~\ref{fig:collimator1} and~\ref{fig:collimator2}, and
a picture of the collimators is given in 
Fig.~\ref{fig:collimators_small.ps}.
In the 2006 ESA test beam experiment the wakefield kick on the beam from
the collimators was studied. The collimators, grouped in two 
sandwiches, were placed into the beam. The kick of the 
beam due to wakesfields has been measured with BPMs placed 
before and after the collimators.

\clearpage
\subsection{Schematic View of Collimator Test Beam Experiment}
\begin{figure}[htbp]
\begin{center}
\includegraphics[width=0.65\textwidth]{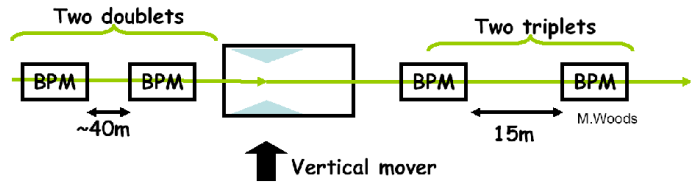}
\end{center}
\vspace*{-5mm}
\caption{Experimental setup at ESA for collimator wakefield measurements.
The collimator jaws are placed in the mover box, as indicated by the flash.
\label{fig:exp_setup}}
\vspace*{-2mm}
\end{figure}

\begin{figure}[htbp]
\begin{center}
\includegraphics[width=0.65\textwidth]{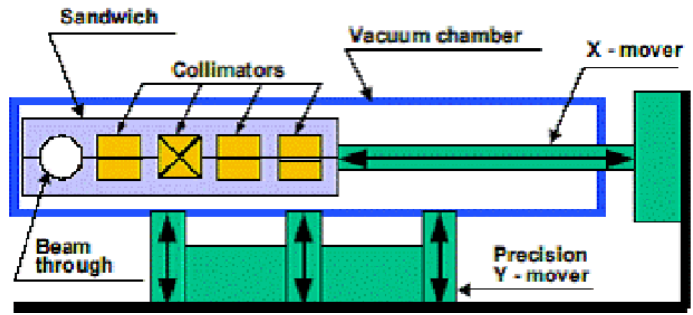}
\end{center}
\vspace*{-5mm}
\caption{Schematic view of the mover box in the experimental setup 
at ESA for collimator wakefield measurements.
\label{fig:exp_setup_sideview}}
\end{figure}

\begin{figure}[htbp]
\vspace*{-1mm}
\begin{center}
\includegraphics[width=0.6\textwidth]{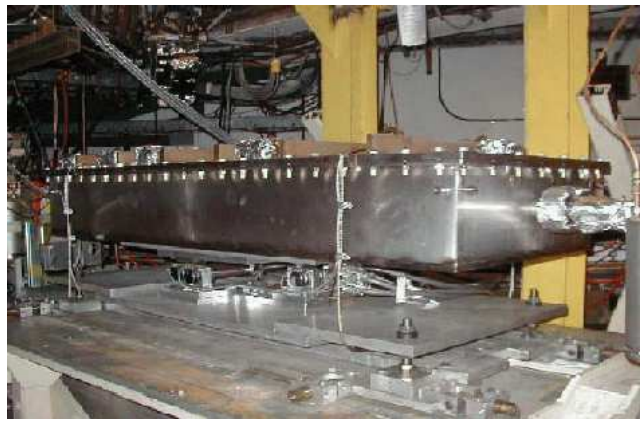}
\end{center}
\vspace*{-5mm}
\caption{Mover box at ESA for measurements of beam kick factors from
wakefield measurements induced by different collimator designs.
\label{fig:mover_box}}
\vspace*{-13mm}
\end{figure}

The schematic view of the collimator test beam setup is shown in 
Figs.~\ref{fig:exp_setup} and~\ref{fig:exp_setup_sideview}.
A detailed description of the beam optics is given in Ref.~\cite{optics2006}.
A picture of the vacuum box containing one collimator sandwich
with four collimators is shown in Fig.~\ref{fig:mover_box}. 
The mover box allows the different collimators to be pushed into the beam 
with precision alignment. As a reference, the mover box also contains
one empty position where the beam can pass through without a collimator.

\clearpage
\subsection{Preliminary Results from ESA Test Beam Runs}
\vspace*{-1mm}

\begin{figure}[htbp]
\vspace*{-4mm}
\begin{center}
\includegraphics[width=\textwidth]{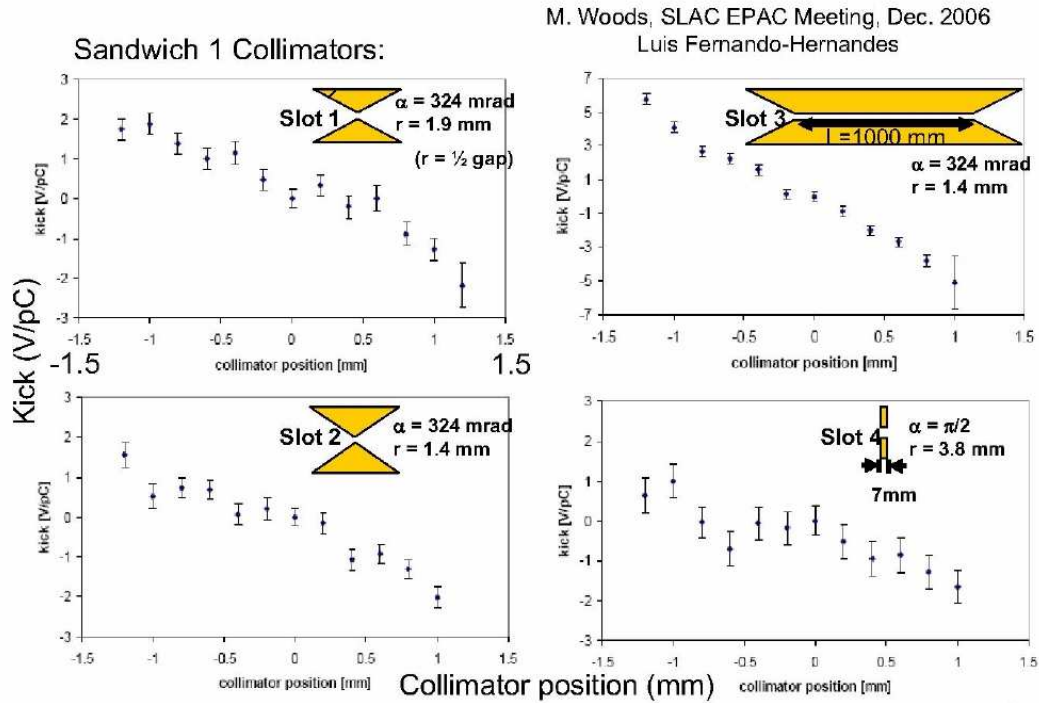}
\end{center}
\vspace*{-7mm}
\caption{Preliminary Results from collimator wakefield measurements at ESA 
in 2006.
The different collimator shapes and sizes lead to different beam kicks 
as a function of the collimator position.
\label{fig:dec06results}}
\vspace*{-5mm}
\end{figure}

The test beam experiments in 2006 showed the expected effect of the 
wakefields on the beam deflection. The size of the deflection 
depends on the position of the beam with respect to the center 
of the collimator pairs. A beam passing through the center of the
collimators pairs is not deflected.
The test beam results are summarized in Fig.~\ref{fig:dec06results} 
for four collimator designs positioned in sandwich 1.
The variation of the beam kick factors on the collimator position, and
the collimator design is demonstrated. The experiment confirms that 
the collimator in sandwich 1 slot 4 has the smallest kick factor compared
to the other collimators, as it is expected from the larger aperture.

\vspace*{-0.5mm}
\subsection{Towards a Comparison of Simulation and Measurements}
\vspace*{-1mm}

\begin{figure}[htbp]
\begin{center}
\vspace*{-10mm}
\includegraphics[width=0.8\textwidth]{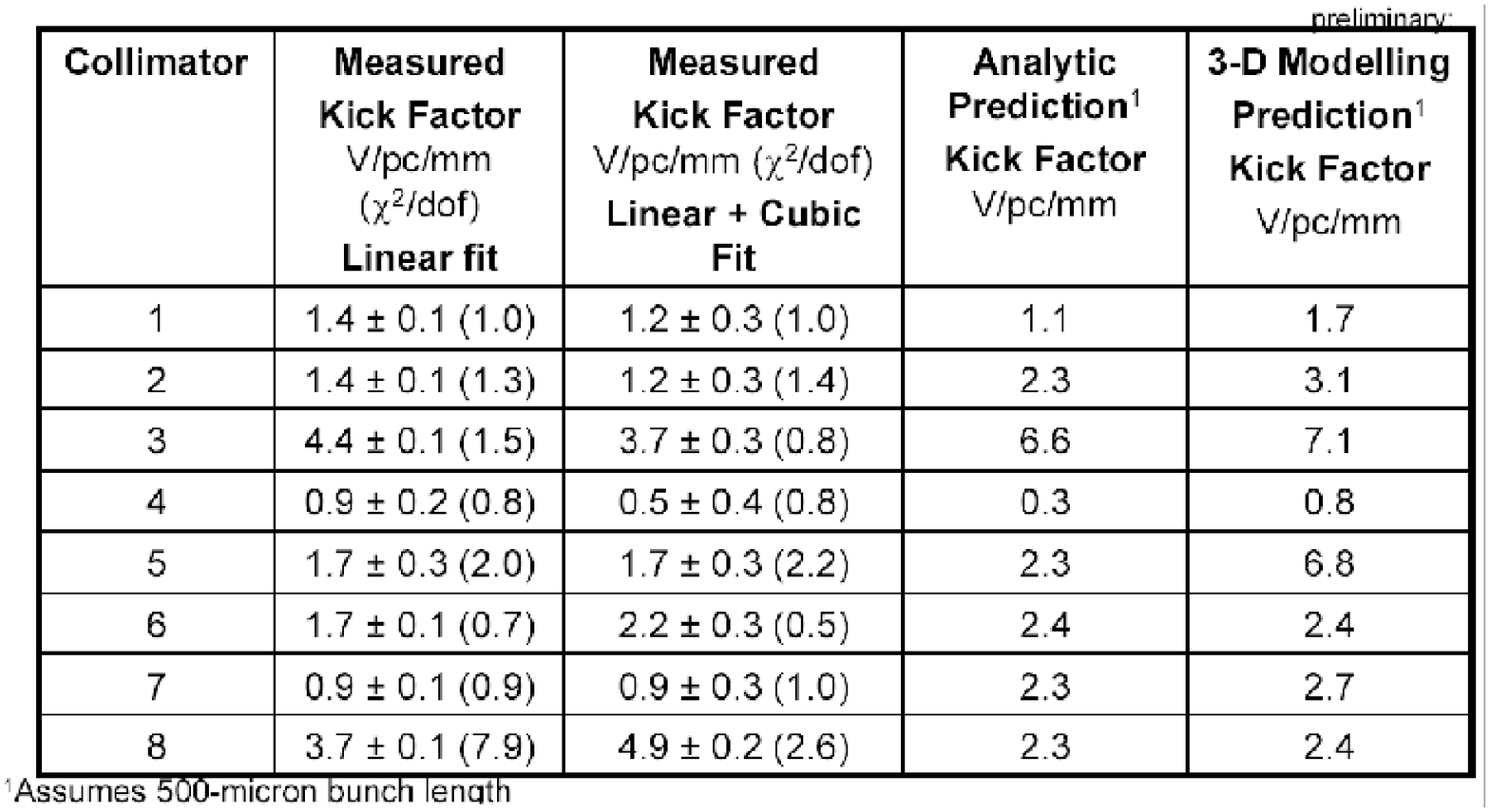}
\end{center}
\vspace*{-5mm}
\caption{Preliminary results from collimator wakefield measurements at ESA
         in 2006.
         Some differences between simulation and measurement
         are still larger than the goal of 10\%. New BPM calibrations and
         refined simulations are under study.
\label{fig:comparison_model2}}
\end{figure}

Precision simulations of the wakefield effects are very challenging
and the test beam results are particularly important for
comparisons with the modelling.
The goal is a determination of kick factors with less than 10\% 
uncertainty. Preliminary results towards a detailed comparison of 
test beam data and simulations are summarized in 
Fig.~\ref{fig:comparison_model2}.

\subsection{2007 Measurements at ESA}
\vspace*{-0.5mm}

Based on first results from 2006 ESA measurements a new 
set of collimators were produced for 2007 measurements and 
applied in two test beam runs.
The new set of collimators explore the following aspects:
shorter collimators, surface roughness, shallow tapers, 
non-linear tapers (exponential).

One collimator of the 2007 test beam studies is identical with that 
used in the 2006 runs in order to study systematic uncertainties in the
measurement precision.
Two test beam runs at the SLAC ESA took place in 2007, one in April and
one in July. For these runs the beam optics is described 
in Ref.~\cite{jackson}.

\section{LHC Collimator R\&D (R. Assmann et al)}

\begin{figure}[htbp]
\vspace*{-5.5mm}
\begin{center}
\includegraphics[width=\textwidth]{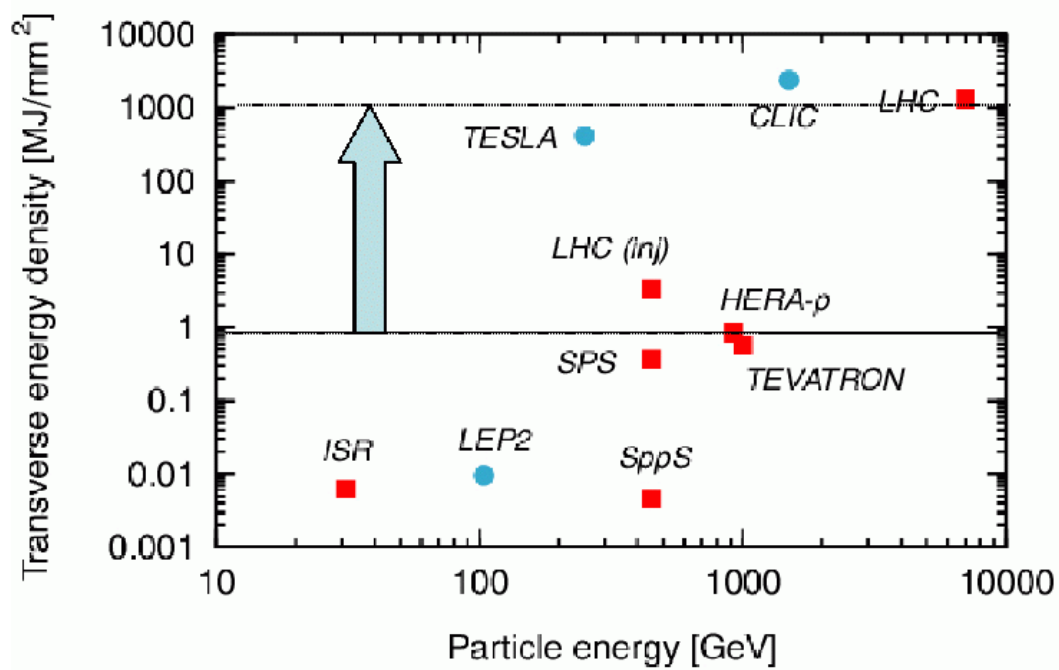}
\end{center}
\vspace*{-5mm}
\caption{Energy density versus particle energy for collimator 
studies\protect~\cite{Assmann:2006bd}. Note the very large energy
density for the LHC.
\label{fig:lhc_energy}}
\vspace*{4mm}
\end{figure}

\begin{figure}[htbp]
\begin{center}
\includegraphics[width=\textwidth]{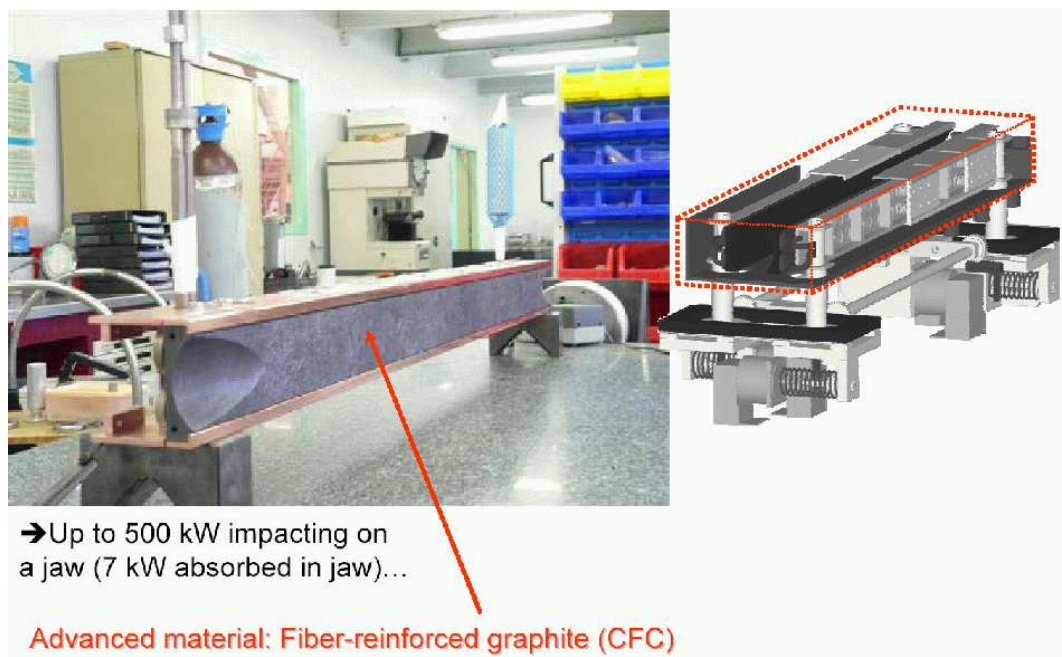}
\end{center}
\vspace*{-5mm}
\caption{LHC collimator test setup at CERN and schematic 
view of high precision mechanical collimator structure.
\label{fig:lhc_collimator}}
\vspace*{-1mm}
\end{figure}

Test beam studies were performed between 2004 and 2006 with the 
SPS beam similar to the LHC beam structure 
($7\mu$s pulse)~\cite{Assmann:2006bd}.  
The energy density with 2MJ/mm$^2$ is about 0.5\% of the LHC beam. 
In particular, survival of shock impact tests have been performed.
Figure~\ref{fig:lhc_energy} (from~\cite{Assmann:2006bd}) shows the
energy density versus the particle energy for the collimator studies.
Up to 500 kW impact on a jaw and 7 kW absorbed is expected.
The LHC collimator is shown in Fig.~\ref{fig:lhc_collimator}.

\section{Some Future Collimator Activities}
\vspace*{-1mm}

For the LHC Phase II, collimator studies continue to play an important
role in the accelerator design. A new test stand at CERN will be 
possible in 2009 which allows to study larger luminosities. 
Collaboration with SLAC through the US LHC Accelerator Research 
Program (LARP) is ongoing. Further aspects of future collimator
test facilities are discussed in the EU Framework 7 proposal
EuCARD (European Coordination for Accelerator Research and 
Development)~\cite{eucard}.
They focus on material damage studies and high density protons beams.
Phase II collimator development is addressed in the work package
ColMat (Collimators \& Materials for higher beam power beam)
and the test facility described 
in the work package HiRadMat (High Radiation Material).
The planned start of EuCARD is January 2009.
Furthermore, the GADGET (Generation And Diagnostics Gear 
for tiny EmiTtance) is discussed within the EU Framework 7 project 
preparation. 
Regarding collimators for future wakefield test beam designs
the BPM resolution and calibration, and their locations are important 
aspects. Furthermore, the possibility of a precise measurement of the 
bunch length is important for a detailed comparison of wakefield 
measurements and simulations. 

\clearpage
\section{\Large Conclusions }
\vspace*{-2mm}

Test beam facilities are very important for the collimator development,
both at the ILC and LHC.
From detailed measurements at the SLAC Endstation A test beam facility 
the influence of collimators on the beam (kick factors for beams off-axis)
have been measured. These measurements guide simulations and material 
design studies.
For collimator design studies the numerical calculations are very 
complex due to the large collimator size compared to the small 
bunch size.
The impact of the wakefields on the beam is an important design 
consideration for the collimators.
In addition to the wakefield aspect other collimator design aspects 
are important, like the beam optics, the interaction region layout, 
design and shaping of the masking, and the inner radius of the 
vertex detector.

At the LHC the large energy densities are a challenge for the collimator
design. The focus is laid on the survivability (machine protection) and
only to a lesser extent on the wakefield aspect.
Test beam facilities remain an important aspect in the future 
for the verification and tuning of the required simulations and 
material studies.

\vspace*{-1mm}
\section*{Acknowledgements} 
\vspace*{-2mm}
I would like to thank my colleagues in the 
collimator project for many fruitful discussions, in particular
Ralph Assmann,
Nigel Watson,
Mike Woods and
Volker Ziemann.
I would also like to thank the organizers of IDTB07 for their support, 
and Nick Walker for comments on the manuscript.
This work is supported in part by the EC under the FP6 Research Infrastructure 
Action - Structuring the European Research Area EUROTeV DS
Project Contract no.011899 RIDS.

\end{document}